\documentclass{article}
\usepackage{authblk}
\usepackage[utf8]{inputenc}
\usepackage{caption}
\usepackage{todonotes}
\usepackage[a4paper, total={6.5in, 10in}]{geometry}
\usepackage{lineno}
\usepackage{setspace}
\usepackage{verbatim}
\usepackage[numbers,sort&compress]{natbib}
\usepackage{multirow}
\usepackage{array}
\usepackage{changepage}
\title{\textbf{How weaponizing disinformation can bring down a city's power grid}}

\author[a]{Gururaghav Raman}
\author[b,$\dagger$]{Bedoor AlShebli}
\author[b,$\dagger$]{Marcin Waniek}
\author[b,*]{Talal Rahwan}
\author[a,*]{Jimmy Chih-Hsien Peng}
\affil[a]{Department of Electrical and Computer Engineering, National University of Singapore, Singapore}
\affil[b]{Computer Science, New York University, Abu Dhabi, United Arab Emirates}
\affil[*]{\footnotesize Joint corresponding authors. E-mail:\ talal.rahwan@nyu.edu;\ jpeng@nus.edu.sg}
\affil[$\dagger$]{\footnotesize Equally contributing authors}

\date{}

\begin{document}
\maketitle

\section*{} 
\begin{adjustwidth}{1.6cm}{1.6cm}
\hspace*{5.8cm}\fontsize{12}{12}\selectfont{\textbf{Abstract}}\vspace*{0.2cm} \\
Social technologies have made it possible to propagate disinformation and manipulate the masses at an unprecedented scale. This is particularly alarming from a security perspective, as humans have proven to be the weakest link when protecting critical infrastructure in general, and the power grid in particular. Here, we consider an attack in which an adversary attempts to manipulate the behavior of energy consumers by sending fake discount notifications encouraging them to shift their consumption into the peak-demand period. We conduct surveys to assess the propensity of people to follow-through on such notifications and forward them to their friends. This allows us to model how the disinformation propagates through social networks. Finally, using Greater London as a case study, we show that disinformation can indeed be used to orchestrate an attack wherein unwitting consumers synchronize their energy-usage patterns, resulting in blackouts on a city-scale.
These findings demonstrate that in an era when disinformation can be weaponized, system vulnerabilities arise not only from the hardware and software of critical infrastructure, but also from the behavior of the consumers.
\end{adjustwidth}

\medskip

\section*{Introduction}
Social technologies have dramatically altered the ways in which conflicts are fought. By allowing belligerents to command the public narrative, these technologies have created a paradigm wherein the most viral information can influence the outcome of wars~\cite{singer2018likewar}. This phenomenon has been exacerbated by social media algorithms that value virality over veracity~\cite{vosoughi2018spread, qiu2017limited}. Unsurprisingly, many notable skirmishes in recent years have used disinformation to manipulate peoples' behavior~\cite{singer2018likewar, harari201821}. Such campaigns have become particularly effective due to the ever-increasing prevalence of big data and machine learning techniques that allow the behavioral patterns of the masses to be analyzed with unprecedented precision. Among the clearest manifestations of such campaigns are the alleged Russian interference into the 2016 US presidential election and the Brexit referendum~\cite{persily20172016,emerson2017model}. These incidents suggest that the microtargeting capabilities provided by companies such as Cambridge Analytica~\cite{nyt2018cambridgeanalytica} can indeed be \textit{weaponized}~\cite{SenateTestimony} to influence the long-term decisions of a society. While many studies have analyzed campaigns targeting long-term social behavior manipulation~\cite{vosoughi2018spread, pennycook2019fighting, del2016spreading, grinberg2019fake, scheufele2019science}, little attention has been given to targeted attacks that use disinformation as a weapon to manipulate social behavior within a limited time span. 

One particularly sensitive target that is vulnerable to behavioral manipulation is critical infrastructure, the attack of which may have drastic implications nationwide. For instance, despite high levels of security, human operators proved to be the weakest link during the Stuxnet attack on the Iranian nuclear program, unwittingly introducing malware into the facilities~\cite{falliere2011w32, nourian2015systems}. Another attack of this kind that drew concern from governments worldwide was the Ukrainian power grid cyberattack of 2015~\cite{Russuia_US_PowerGrid1, Russuia_US_PowerGrid2}. In this incident, attackers deliberately cut off the power supply for 230,000 residents for several hours using operator credentials harvested through one particular form of disinformation, namely, spear-phishing~\cite{UkraineAttack}. 

In this study we focus on the power grid---a choice motivated by the devastation caused by historical power outages including human casualties and massive financial losses~\cite{CostOfBlackout_NY,NorthEastBlackout_Wiki, VenezuelaBlackout}. Yet, while numerous blackout prevention and mitigation strategies have been proposed in the literature~\cite{slowCoherency, ExposeHiddenFailures, MASDistRestoration, van2011vehicle, MASPwrRestoration, SelfHealingInfra, RAcascading, you2003self, ReliabilityPerspective, SectionalizationMicrogrids, NetMicrogridsSelfHealing, nagarajan2016optimal}, the link between disinformation and blackouts has never been studied to date. Driven by this observation, we seek to answer the following question: can an adversary bring down a city's power grid using disinformation without any physical or cyber intrusions?

\section*{Attack propagation through social networks}
\begin{figure}[t]
	\centering
	\includegraphics[width=\columnwidth]{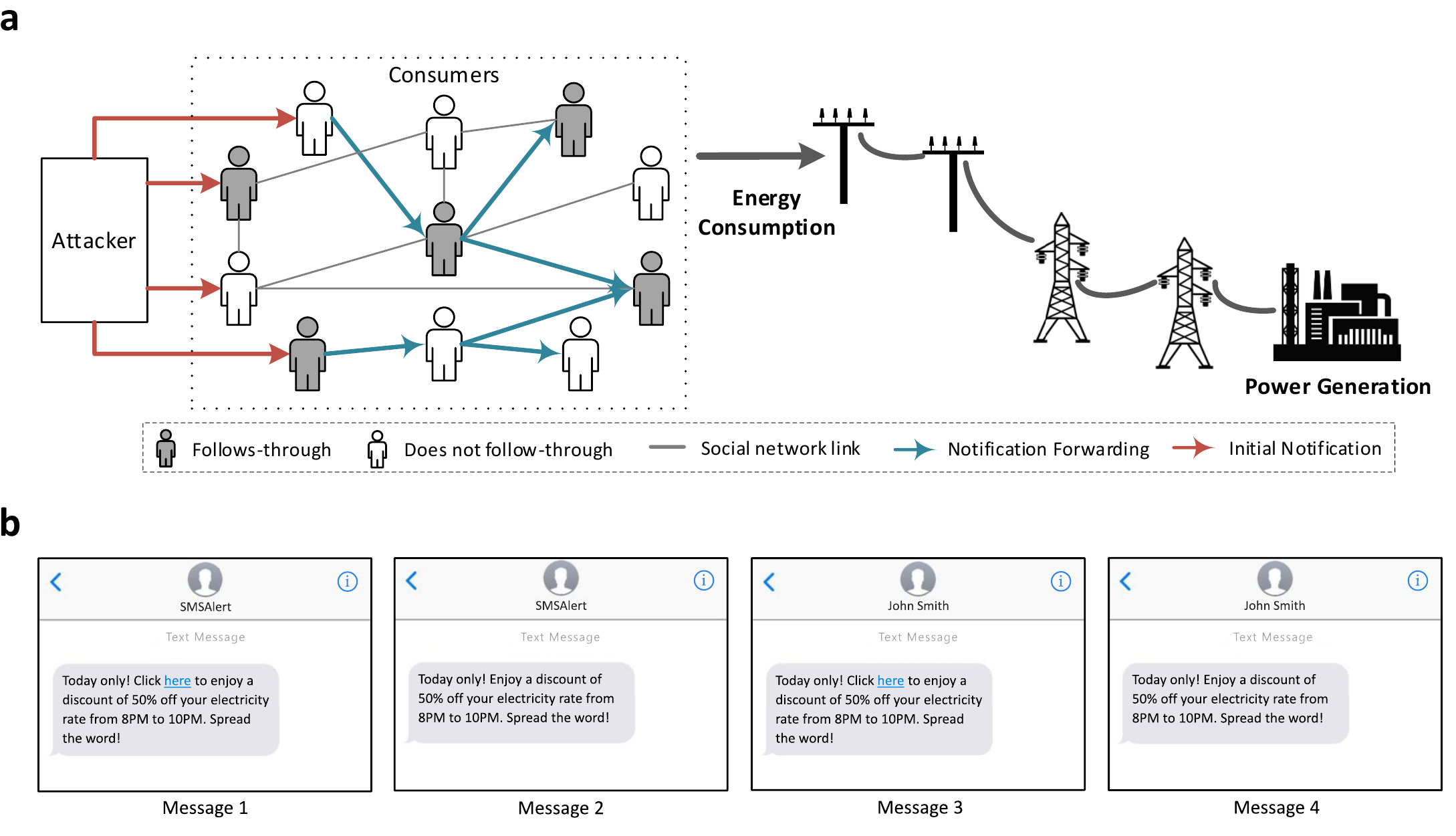}
	\caption{\textbf{An overview of a disinformation attack on the power system.} \textbf{a}, Illustrating how the disinformation attack is launched from a single attacker and amplified via propagation in social networks, thereby altering the energy consumption patterns of a portion of the population. Importantly, not every recipient follows-through, nor does a recipient have to follow-through in order to forward the notification. \textbf{b}, The disinformation notification shown to participants in different conditions, which vary depending on whether or not the notifications contain an external link, and whether the sender is a stranger (assumed to be the attacker who uses spoofing services to mask the sender as \textit{SMSAlert}) or a friend (named \textit{John Smith} in the survey). }
	\label{Infographic}
\end{figure}

We consider an attack in which an adversary attempts to manipulate the behavior of citizens by sending fake discount notifications encouraging them to shift their energy consumption into the peak-demand period. Such a shift may result in the tripping of overloaded power lines, leading to blackouts (see Methods).
Ultimately, the success of such an attack depends on the \textit{follow-through rate}, i.e., the fraction of people who behave as intended by the attacker. Here, the social aspect plays a crucial role, since people may unknowingly amplify the attack by forwarding the disinformation notification to their friends, who in turn forward it to their own friends, and so on; see Fig.~\ref{Infographic}a for an illustration. (Note that the term ``friend'' is borrowed from the context of social media to refer to any ``acquaintance''.) To model the spread of disinformation in the social network, we use two standard models of influence propagation, namely, independent cascade~\cite{goldenberg2001using}, and linear threshold~\cite{kempe2003maximizing}. While a person may receive the same notification from more than one friend, the main difference between the two models lies in the way in which that person is influenced by such repeated exposure to the notification. In the independent cascade model, every exposure has an independent probability to persuade the recipient to modify their behavior---these probabilities constitute the main parameters of the model. In contrast, each person in the linear threshold model has a ``threshold'' specifying the number of exposures required for them to modify their behavior---these thresholds constitute the main parameters of the model; see Methods for more details.

In our simulations, the values of these parameters are determined based on the responses of 5,124 surveyed participants who were recruited through Amazon Mechanical Turk. Specifically, the participants were shown a message notifying them of a discount of 50\% in their electricity rate from 8PM to 10PM. They were then asked to specify the likelihood of them changing their electricity-use patterns to take advantage of this discount, and the likelihood of them forwarding this message to their friends. We tested two factors that may influence the behavior of the participants: (i) the notification sender, and (ii) the notification content. As for the first factor, while such notifications are typically received from the power utility, we analyzed the cases when they are instead received from either a stranger or a friend. We considered these two possibilities since some people may receive the spoofed message directly from the attacker (who is a stranger to them), while others may receive it indirectly through friends who forward it to them. As for the second factor---the notification content---we analyzed two variants: one where the discount can only be availed by clicking on an external link, and another where the discount is unconditional. This manipulation allows us to understand the differences, if any, between the context of phishing and spam attacks---which require the recipients to click on an external link embedded in the message---and the context of our disinformation attack---where no such link is necessary. Accordingly, the participants were randomly assigned to one of four conditions: (i) receive a notification with a link from a stranger; (ii) receive a notification without a link from a stranger; (iii) receive a notification with a link from a friend; (iv) receive a notification without a link from a friend. The corresponding messages that were displayed to participants are depicted in Fig.~\ref{Infographic}b. They were then asked questions to determine how they would react to these messages. Here, participants were further split into two groups depending on the influence model being studied, since the parameters of each model require the questions to be framed differently. The complete survey along with a summary of the results is provided in Supplementary Note~1. 

\begin{figure}[t]
\centering
\includegraphics[width=\columnwidth]{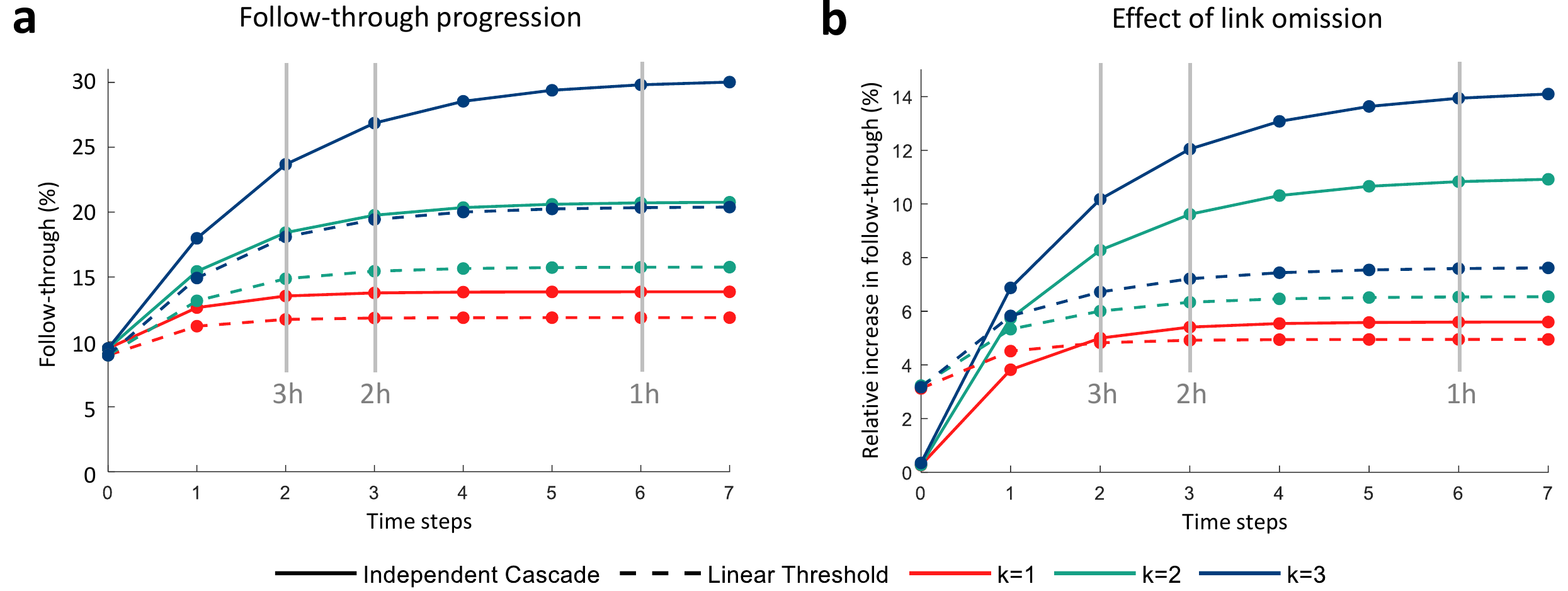}
\caption{\textbf{Attack diffusion.} Analyzing the follow-through rates when the attacker initially sends the notification to 200,000 individuals in scale-free networks consisting of 1 million individuals, given two influence propagation models (independent cascade and linear threshold) and three values of $k$ (representing the number of friends to whom each recipient considers forwarding the notification). The vertical lines indicate the number of people who follow-through at the peak energy demand period when each time step takes 1h, 2h, and 3h, assuming that the attack is launched 6 hours prior to this period. \textbf{a}, The average number of people who follow-through after each time step of the spread of disinformation. \textbf{b}, The relative increase in the number of people who follow-through due to the omission of the link from the notification.}
\label{Diffusion_results}
\end{figure}

Next, the survey outcomes were used to evaluate the extent to which the attack is amplified when the notification is propagated in a social network. To this end, we conducted simulations on 100 randomly generated scale-free networks~\cite{barabasi1999emergence}, each consisting of 1 million nodes, with an average degree of 10. For each node, its propensity to follow-through and forward the notification is set to match that of a randomly-chosen survey participant.
Within our simulations, nodes that receive the notification forward it to $k$ of their friends with a probability that corresponds to their respective preferences. For instance, if $k=3$ and a participant specifies their likelihood of forwarding the message to be 50\%, then the corresponding node chooses 3 friends at random, and forwards the notification to each of them with a probability of 50\%.
Every simulation proceeds in time steps as follows. The nodes that receive the notification in a time step $t$ may decide to follow-through (depending on their preferences), and may forward it to their friends (i.e., the other nodes that are connected to them in the network) who would then receive it in time step~$t+1$. This process continues until the state of the network remains unchanged in two consecutive time steps.
Given the two influence propagation models and $k\in\{1,2,3\}$, Fig.~\ref{Diffusion_results}a depicts the average number of people who follow-through in each time step, assuming that the initial notification is sent to 20\% of the individuals in the social network. 
The vertical lines in the figure indicate the follow-through rates at the peak demand period when each time step takes 1h, 2h, and 3h, assuming that the initial notification is sent 6h before this period.
As can be seen, even if each time step takes 3 hours, the percentage of people following-through can reach as high as 23.7\%, and no less than 11.7\% according to the most conservative estimate. Recall that unlike the case of phishing and spam attacks, the disinformation attack considered here does not require an external link. To evaluate how this difference affects the impact of the attack, we run similar simulations based on the responses of participants who were shown a message containing an external link. We found that the omission of the link always increases the follow-through rate. For instance, even if each time step takes 3h, the increment ranges from 4.8\% to 10.2\%; see Fig.~\ref{Diffusion_results}b. As shown in Supplementary Note~2, similar results were obtained when the initial notification was sent to 10\% and 30\% of the individuals in the network. 

\section*{Attack impact on the power grid}
\begin{figure}[t] 
	\centering
	\includegraphics[width=\columnwidth]{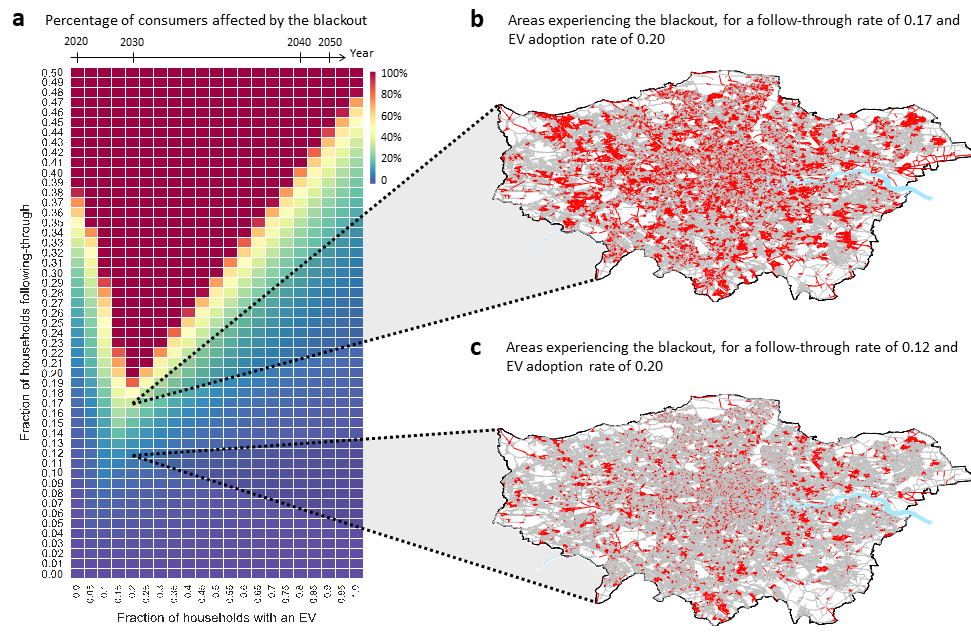}
	\caption{\textbf{Impact of an attack on the power distribution network of Greater London}. \textbf{a}, The percentage of consumers suffering from a blackout as a result of the attack given different follow-through rates and EV adoption rates. The figure also highlights the columns corresponding to projected EV adoption rates for the UK in the years 2020, 2030, 2040, and 2050. \textbf{b}, Visualization of the status of every power distribution line in the system for a follow-through and EV adoption rates of 0.17 and 0.20, respectively. Grey indicates active lines, whereas red indicates lines that have tripped as a result of overloading. \textbf{c}, The same as (\textbf{b}), but for follow-through and EV adoption rates of 0.12 and 0.20, respectively.}
	\label{LondonMaps}
\end{figure}

Having analyzed the number of people who follow-through on the notification, we now analyze the impact caused by such behavior manipulation on the power grid. To this end, we modelled the power grid of Greater London (see Supplementary Note~3) and simulated the behavior of residential energy consumers. Importantly, our model considers residential electric vehicle (EV) adoption since the owners of such EVs control a substantial amount of deferrable energy, and thus can cause greater harm when manipulated by an adversary. We vary the EV adoption level in the city, and model the capacity upgrades that are necessary for the grid to support the demand corresponding to each such level \cite{quiros2018electric, muratori2018impact, coignard2019will}. Fig.~\ref{LondonMaps}a presents the percentage of consumers who experience a blackout given varying follow-through and EV adoption rates. As can be seen, increasing the EV adoption up to 20\% increases the system vulnerability to the attack, whereas beyond 20\%, the system resilience increases, i.e., it requires a greater follow-through rate to achieve the same attack magnitude. This trend is caused by two opposing forces: (i) increased vulnerability due to the consumers controlling more deferrable energy, and (ii) increased resilience due to the grid's upgraded capacity. When the EV adoption is smaller than or equal to 20\%, the former force outweighs the latter, and hence we see an increase in the system vulnerability. The opposite is true when the EV adoption exceeds 20\%, leading to the observed increase in resilience.

Now, consider the case when the EV adoption is 20\% and the attack is launched 6 hours prior to the peak demand period. Assuming that every time step in the influence propagation takes 1h, the follow-through rate ranges from 11\% to 30\% (see Fig.~\ref{Diffusion_results}a), resulting in a blackout for 7.3\% to 100\% of the residents, respectively (see Fig.~\ref{LondonMaps}a). To put it differently, behavioral manipulation through disinformation can lead to a full blackout at the city-scale. Next, to get a sense of the distribution of the blackout across the city, we depict the state of the system corresponding to two different cells in the heat map; see Fig.~\ref{LondonMaps}b and \ref{LondonMaps}c. As can be observed, the impact is dispersed throughout the city rather than being concentrated in very few massive pockets. Note that the results shown thus far are for the case where, for any given EV adoption rate, the grid is assumed to be upgraded to support exactly that rate. However, if the grid is upgraded to support more than this rate, the impact of the attack will be substantially alleviated. Taking the year 2025 as an example, if by then the grid was not upgraded since 2020, then a mere 5\% follow-through rate can bring the grid down completely. On the other hand, if the grid in 2025 was upgraded to support the projected EV adoption until 2030, then even a 100\% follow-through rate would cause a blackout for less than 20\% of the residents. These results highlight the need for future grid upgrades to not only be dictated by the technical aspects governed by physical laws, but also consider the behavioral aspects of the consumers who may act unpredictably and irrationally, especially when subjected to disinformation. However, since grid upgrades come at a high cost to the power utility~\cite{coignard2019will}, perhaps a more realistic solution would be to focus on increasing the awareness of the consumers and immunizing them against disinformation.

\section*{Discussion}
While the literature on power grids focuses on the advantages of increasing the active consumer engagement and coordinating their consumption patterns~\cite{nicolson2017tailored, tiefenbeck2019real,muratori2018impact, coignard2019will}, we demonstrated that such engagement makes the grid more vulnerable to behavior manipulation attacks. In particular, we showed how an adversary can use disinformation to manipulate the behavior of energy consumers by sending them fake notifications that encourage them to shift their energy usage into the peak demand period. Our surveys showed that people are willing to not only follow-through on such notifications, but also forward them to their friends, thereby amplifying the attack. This is partly attributed to the fact that such notifications, unlike those used in spam and phishing attacks, do not include any external links in the message. We quantified the impact of such an attack on a city-scale, taking Greater London as an example and considering the projected EV adoption levels in the city. This demonstrated that an adversary can indeed cause blackouts throughout the city, without tampering with the hardware or hacking into the control systems of the power grid, but rather focusing entirely on behavior manipulation. On a broader note, our study is the first to demonstrate that in an era when disinformation can be weaponized, system vulnerabilities arise not only from the hardware and software of critical infrastructure, but also from the behavior of the consumers.

\section*{Methods}
\medskip
\noindent \textbf{Disinformation content.}
The lines in the power grid have limited capacity that cannot easily be expanded due to financial constraints. Consequently, legacy networks are often unable to support extremely unlikely scenarios, e.g., when the peak demand increases abruptly in a totally unexpected manner as in the scenario considered in our paper. This limitation is further exacerbated by the inclusion of EVs, which have the highest real power consumption compared to household appliances. Given this limitation, an adversary may engineer a scenario aimed at breaching the capacity limits of the power lines. 
One way to achieve this goal is to persuade as many people as possible to shift their energy consumption into the peak demand period, when the grid is already at its most vulnerable state. In our setting, this is achieved by spreading a fake discount message informing people of a discount effective during the peak period.

\medskip
\noindent \textbf{Models of influence.}
We use two fundamental models of influence propagation in social networks, namely, independent cascade~\cite{goldenberg2001using} and linear threshold~\cite{kempe2003maximizing}. Both models start with an ``active'' subset of nodes, called the \emph{seed set}. Then, the influence of these active nodes propagates through the network in time steps. Formally, let $V$ denote the set of nodes in the network, and $N_v$ denote the neighbors of node $v \in V$. Moreover, let $A(t)\subseteq V$ be the set of active nodes at time step $t$, implying that $A(1)$ is the seed set. The mechanism of influence propagation depends on the model being used. In the independent cascade model, every pair of nodes has an activation probability, $p : V \times V \rightarrow [0,1]$. Then, in each time step $t>1$, every node $v$ that became active at time step $t-1$ activates every inactive neighbor $w\in N_v\setminus A(t-1)$ with probability $p(v,w)$. The propagation terminates when $A(t)=A(t-1)$. On the other hand, in the linear threshold model, every node $v \in V$ is assigned a threshold, $k_v$, such that $0\leq k_v\leq|N_v|$. Then, in each time step $t>1$, every node $v\in V\setminus A(t-1)$ becomes active if the following holds: $|A(t-1) \cap N_v| \geq k_{v}$. Again, the propagation terminates when $A(t)=A(t-1)$.

In our study, we introduced three modifications to the influence propagation models in order to reflect the attack scenario being considered. First, we decoupled the state of being activated from the state of influencing others. Specifically, being activated in our setting means deciding to follow-through on the notification being received. In contrast, influencing neighbors means forwarding the notification to one's friends. As such, one may be activated without necessarily influencing others, and vice versa. 
The second modification involved distinguishing between those who receive the notification from a stranger (who is the attacker in our case) and those who receive it from their friends (who forwarded the notification to them). This distinction matters, since the way in which recipients react to the notification is affected by the identity of the sender, as evident from the outcome of our surveys (see Supplementary Note~1).
The third modification involved allowing the individuals to influence only a subset of their friends. This makes the model more realistic, since people are usually not restricted to forwarding a message to either \emph{all} or \emph{none} of their friends.
It should be noted that for an individual to be included in the seed set, it is not sufficient for them to simply receive the notification from the stranger; they have to also decide to forward it to their friends. This is especially important in the linear threshold model, since our definition of the seed set means that the model cannot be parameterized solely based on thresholds; it also needs parameters specifying the likelihood of the nodes to forward a notification received from a stranger.

The influence propagation models used in our simulations were parameterized based on the survey outcomes. In particular, for the independent cascade model, each participant specified (i) their likelihood to follow-through, and (ii) their likelihood to forward the notification, when the participant received it either from a stranger or a friend (note that an individual may first receive the notification from the stranger, and then again from a friend at a later time step). On the other hand, for the linear threshold model, participants specified (i) their likelihood to follow-through and forward the notification when received from a stranger, and (ii) the number of friends that they need to receive the notification from in order for them to follow-through and forward it to others (see Supplementary Note~1). Participants specified their likelihoods on a Likert scale from 0 to 10. Each response, $x \in[0,10]$, was then converted to the probability $\frac{x}{10}$. Finally, in our simulations, the number of friends that each individual considered forwarding the notification to was determined based on a parameter $k\in\{1,2,3\}$.

\medskip
\noindent \textbf{Power grid specifications.} 
Due to its sensitivity, comprehensive data describing the power distribution network of Greater London is not publicly available. We therefore built our own model based on data obtained from OpenStreetMap~\cite{OpenStreetMap}, under the reasonable assumption that power lines (i.e., overhead lines or cables) are laid alongside roads. In particular, we start by extracting the road network and the location of every building therein. Next, we obtain the locations of the 9 substations that feed Greater London~\cite{NationalGridUK}.
Note that each building is electrically connected to a single substation, which is typically the one closest to it. With this in mind, we divide the road network of Greater London into 9 subnetworks, one per substation, and construct 9 spanning trees connecting the buildings within each subnetwork to the corresponding substation. We construct these spanning trees using a modified version of Kruskal's algorithm~\cite{kruskal1956shortest} while taking into consideration various technical and economical constraints. The edges in these spanning trees represent the power lines in our simulations.
As for the loading of each line in the power grid, we assume that every building represents a household, the energy consumption of which is modelled using statistics obtained from~\cite{richardson2010domestic, quiros2018statistical}. For more details, see Supplementary Notes~3 and 4.  

\medskip
\noindent \textbf{Power simulations.}
Since the energy consumption of a residence depends on its occupancy~\cite{richardson2010domestic}, residences in our simulation were assigned occupancy values based on UK National Statistics~\cite{quiros2018statistical}. As for the EVs, we restrict our attention to residential rather than commercial EVs, since the latter have a negligible effect on the peak demand when compared to the former~\cite{FES_UK_NationalGrid2018}. In each simulation, EV owners and notification recipients were selected randomly based on the EV adoption rate and disinformation propagation rate in the simulation, respectively. Each resident was assigned a daily load profile depending on whether they own an EV and whether they follow-through on the notification (see Supplementary Note~4 for how these load profiles were generated). Subsequently, power flows within the distribution network were calculated. 
Finally, we make the reasonable assumption that the capacity of each line in the distribution network is limited to 10\% over the peak power flow in that line under regular circumstances, i.e., when no resident receives the notification from the attacker.

Finally, we analyze how the distribution network is affected by the attack. In this analysis, among the many variables that could be considered such as voltage and reactive power flows, we focused on line capacity limits since they are the most dominantly affected by energy consumption patterns, and are the most critical to power system stability~\cite{quiros2018electric, coignard2019will}.
Once a line overloads it goes offline, leading all lines below it in the power distribution tree to go offline as well. By averaging over 100 such simulations, we obtained the fraction of residences suffering from a blackout that were depicted in Fig.~\ref{LondonMaps}a. The illustrations in  Fig.~\ref{LondonMaps}b and \ref{LondonMaps}c represent a single simulation each.

\section*{Ethics statement}
The research was approved by the Institutional Review Boards of New York University-Abu Dhabi and the National University of Singapore.

\section*{Data availability} 
All power systems data that were used in this study are available to download from~\cite{website_data}. The codes for generating the specific plots are available from the corresponding authors upon request.

\small
\bibliographystyle{unsrt}
\bibliography{ref}
\normalsize

\end{document}